\documentclass[conference]{IEEEtran}
\IEEEoverridecommandlockouts
\usepackage{cite}
\usepackage{amsmath,amssymb,amsfonts}
\usepackage{algorithmic}
\usepackage[pdftex]{graphicx}
\usepackage{booktabs}
\usepackage{textcomp}
\usepackage{xcolor}
\usepackage{hyperref}
\usepackage{tikz}
\usepackage{multirow}
\usetikzlibrary{arrows.meta}
\usetikzlibrary{shapes.geometric, arrows}
\tikzstyle{startstop} = [rectangle, rounded corners, minimum width=1cm, minimum height=0.5cm,text centered, draw=black, fill=red!30]
\tikzstyle{process1} = [rectangle, minimum width=2cm, minimum height=0.5cm, text centered, draw=black, fill=orange!30]
\tikzstyle{process2} = [rectangle, minimum width=1cm, minimum height=0.5cm, text centered, draw=black, fill=green!30]
\tikzstyle{arrow} = [thick,->,>=stealth]
\newcommand{\RNum}[1]{\uppercase\expandafter{\romannumeral #1\relax}}
\def\BibTeX{{\rm B\kern-.05em{\sc i\kern-.025em b}\kern-.08em
    T\kern-.1667em\lower.7ex\hbox{E}\kern-.125emX}}

\begin{document}

\title{Graphs are everywhere - Psst! In Music Recommendation too\\
}

\author{\IEEEauthorblockN{Bharani Jayakumar}
\textit{Technische Universität Ilmenau}\\
Ilmenau, Germany \\
bharani.jayakumar@tu-ilmenau.de
\and
\IEEEauthorblockN{Orkun Özoğlu}
\textit{Independent Researcher} \\
Bursa, Türkiye \\
oozoglu@pm.me
}

\maketitle

\begin{abstract}
In recent years, graphs have gained prominence across various domains, especially in recommendation systems. Within the realm of music recommendation, graphs play a crucial role in enhancing genre-based recommendations by integrating Mel-Frequency Cepstral Coefficients (MFCC) with advanced graph embeddings. This study explores the efficacy of Graph Convolutional Networks (GCN), GraphSAGE, and Graph Transformer (GT) models in learning embeddings that effectively capture intricate relationships between music items and genres represented within graph structures. Through comprehensive empirical evaluations on diverse real-world music datasets, our findings consistently demonstrate that these graph-based approaches outperform traditional methods that rely solely on MFCC features or collaborative filtering techniques. Specifically, the graph-enhanced models achieve notably higher accuracy in predicting genre-specific preferences and offering relevant music suggestions to users. These results underscore the effectiveness of utilizing graph embeddings to enrich feature representations and exploit latent associations within music data, thereby illustrating their potential to advance the capabilities of personalized and context-aware music recommendation systems.
\end{abstract}


\section{Introduction}
In recent years, the integration of graphs has revolutionized recommendation systems across diverse domains by leveraging their ability to model complex relationships between entities. Within the realm of music recommendation, graphs have emerged as a potent tool for enhancing genre-based recommendations through the augmentation of traditional features like Mel-Frequency Cepstral Coefficients (MFCC) with graph embeddings. This paper addresses the significant challenge of improving music recommendation accuracy through the application of Graph Convolutional Networks (Graph CN), GraphSAGE, and Graph Transformer models. These models facilitate the learning of embeddings that capture nuanced associations between music items and genres encoded in graph structures.

The motivation behind this research stems from the critical need to surpass the limitations of conventional recommendation approaches, which often struggle to effectively capture and utilize contextual relationships inherent in music data, such as genre preferences and artist similarities. By harnessing the rich information encapsulated in graph representations, the aim is to significantly enhance the precision and relevance of music recommendations tailored to individual user preferences. This study empirically evaluates and compares the performance of Graph CN, GraphSAGE, and Graph Transformer models on real-world music datasets, demonstrating their superiority over traditional methods. The findings underscore the efficacy of these graph-based approaches in enriching feature representations and uncovering latent associations within music data, thereby advancing the frontier of personalized and context-aware music recommendation systems.

Integrating graph-based methods into music recommendation systems, however, presents several challenges. These include scaling graph algorithms to handle large-scale music datasets effectively and efficiently incorporating diverse types of music-related information into graph structures. Addressing these challenges is crucial for realizing the full potential of graph-enhanced recommendation systems in delivering more personalized and context-aware music experiences.

The rest of the paper is organized as follows. Section \RNum{2} talks about the related work in the current field of interest. Section \RNum{3} forms the crux of the paper with the realized concept and the implementation of the same explaining the methods. Section \RNum{4} provides insights on the findings based on the implementation while Section \RNum{5} opens the floor for discussion and concludes by summarizing the paper. 

\section{Related Work}
In recent years, recommendation systems have benefited significantly from the integration of graph-based methodologies, which excel in modeling complex relationships between entities. Nam et al. \cite{nam2019deep} employed deep learning techniques for audio-based music classification and tagging, achieving notable advancements in genre recognition and content-based recommendation. Their study underscores the efficacy of deep neural networks in capturing intricate musical features, thereby enhancing recommendation accuracy based on content similarity.

Wang et al. \cite{wang2021graph} reviewed graph learning-based recommendation systems and summarized the importance of graph structures to improve the reliability and accuracy of recommendations provided. This paper also explains in detail how this novel research area involving graphs could be expanded to multiple domains. This motivated our investigation into graph-based embeddings for genre-based music recommendation, emphasizing the importance of capturing contextual dependencies to enhance user satisfaction and engagement.

Sakurai et al. \cite{sakurai2022graph} proposed a framework for playlist generation using graph-based methods especially the usage of Knowledge Graph (KG) and Reinforcement Learning (RL), integrating user preferences and song characteristics encoded within graph structures. Their approach highlights the significance of context-aware recommendation systems in adapting to individual user tastes and preferences by capturing the target users’ long-term preferences. Similarly, recent advancements in graph embedding techniques such as GraphSAGE and Graph Transformer models \cite{hamilton2017inductive, velivckovic2018graph} have shown promise in capturing hierarchical and structural information in graphs, making them suitable for enhancing recommendation systems by learning more informative representations of music items and user preferences.

While existing literature predominantly focuses on either deep learning approaches for content-based recommendation or graph-based methods for collaborative filtering, our research uniquely integrates both aspects. We extend current understanding by systematically comparing GCN \cite{kipf2016graph}, GraphSAGE, and GT models in the context of genre-based music recommendation. Unlike previous studies that often specialize in either content-based or collaborative filtering approaches, our work explores the synergistic benefits of combining graph embeddings with traditional feature extraction methods like Mel-Frequency Cepstral Coefficients (MFCC) \cite{mfcc}, aiming to provide more accurate and personalized music recommendations.

This paper contributes to the existing body of knowledge by empirically evaluating and demonstrating the effectiveness of GCN, GraphSAGE, and GT models on real-world music datasets. Our findings not only showcase superior performance in genre-based music recommendation but also highlight the scalability and adaptability of graph-based methodologies in handling diverse music-related information. By bridging the gap between deep learning and graph theory, this study extends current understanding and provides practical insights for developing more robust and context-aware music recommendation systems.

\section{Methodology}
\subsection{Dataset Description}
The dataset used in this study is balanced, consisting of 1,000 songs for each of the 8 genres: Electronic, Experimental, Folk, Hip-Hop, Instrumental, International, Pop, and Rock. This results in a total of 8,000 songs. The balanced nature of the dataset ensures that the model is trained evenly across all genres. \cite{fma2016}

\subsection{Data Preprocessing}
Each song in the dataset was processed by selecting a random 5-second window, which was then saved as a new audio sample. This step ensures uniformity in the length of audio samples, facilitating consistent feature extraction and model training. Mel-Frequency Cepstral Coefficients (MFCC) features (dimension = 30) were extracted from the 5-second audio samples. MFCCs provide a compact representation of the power spectrum of audio signals, capturing essential characteristics that are pertinent to genre classification.

\subsection{Graph embedding learning}
\begin{figure}[h]
    \centering
    \includegraphics[width=0.45\textwidth]{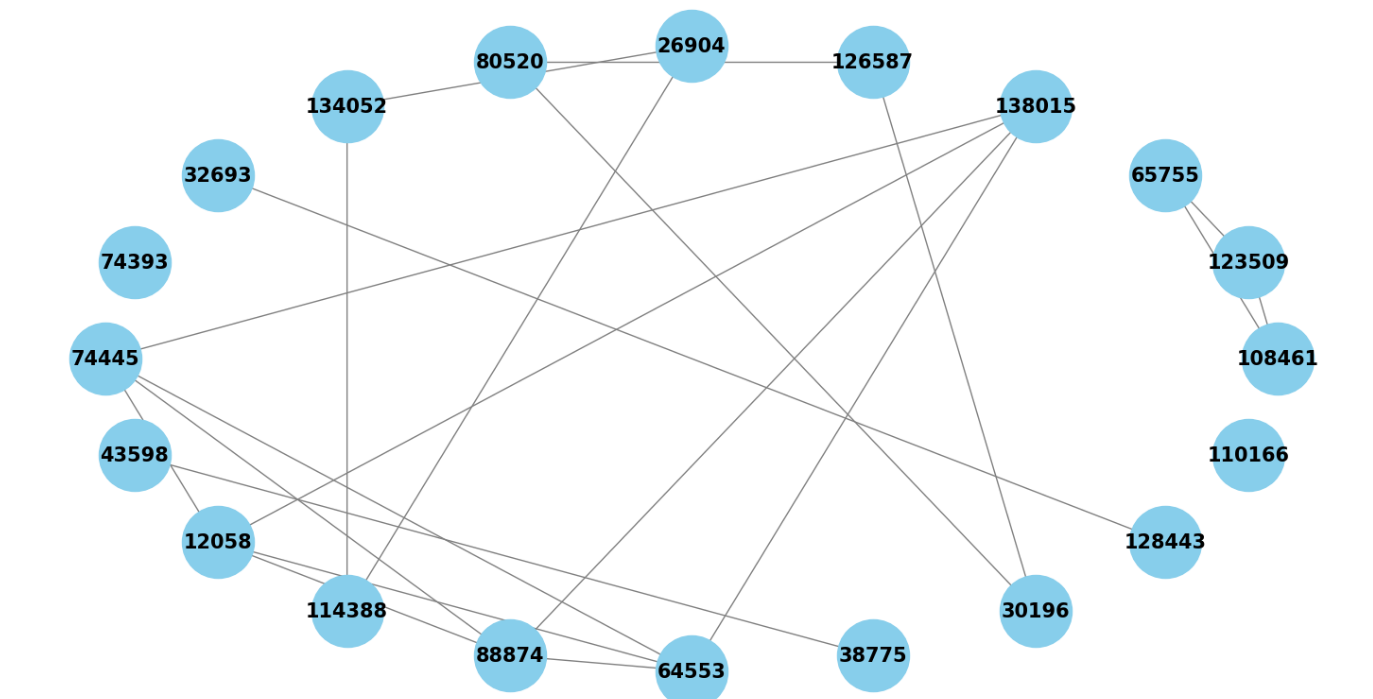}
    \caption[Graph of Songs (20 random songs)]{Graph of Songs (20 random songs) based on Genres. The edges between the songs mean that they belong to the same genre.}
    \label{sample_graph}
\end{figure}

To enrich the MFCC features, the dataset was converted into a graph structure where each song represents a node. An edge exists between two nodes if they belong to the same genre. Figure \ref{sample_graph} gives an intuitive illustration of songs as nodes and connected if they belong to the same genre. Note that the representative figure shown is just for a random selection of 20 songs from the dataset. GraphSAGE and GCN were the two methods used to refine the features of MFCC.

\subsubsection{GraphSAGE}
\begin{figure}[hbt!]
    \begin{minipage}[b]{0.22\textwidth}
        \centering
        \begin{tikzpicture}[node distance=0.1cm, thick, main/.style={circle, draw=black, minimum size=0.25cm}]
        \node[main, fill=purple!50] (central) at (0, 0) {};
        \foreach \i in {1,...,10}
        \node[main, fill=orange!50] (orange\i) at ({90+360/7*\i}:1cm) {};
        \foreach \i in {1,...,10}
        \draw[->, purple!80, thick] (central) -- (orange\i);
        \foreach \i in {10,...,15}
        \node[main, fill=white] (node\i) at ({90+360/10*\i}:1.5cm) {};
      \node[above=0.25cm, purple] at (central) {};
    \end{tikzpicture}
    \end{minipage}
    \hfill
    \begin{minipage}[b]{0.22\textwidth}
        \centering
        \begin{tikzpicture}[node distance=0.1cm, thick, main/.style={circle, draw=black, minimum size=0.25cm}]
        \node[main, fill=green!50] (central) at (0, 0) {};
        \foreach \i in {1,...,10}
        \node[main, fill=orange!50] (orange\i) at ({90+360/7*\i}:1cm) {};
        \foreach \i in {1,...,10}
        \draw[->, dashed, orange!80, thick] (orange\i) -- (central);
        \foreach \i in {10,...,15}
        \node[main, fill=white] (node\i) at ({90+360/10*\i}:1.5cm) {};
      \node[above=0.25cm, purple] at (central) {};
    \end{tikzpicture}
    \end{minipage}
\caption{Illustration of GraphSAGE - Leftside: The graph before sample and aggregation. Rightside: The neighboring nodes of the same characteristics are sampled aggregated and represented}
\label{GraphSAGE}
\end{figure}
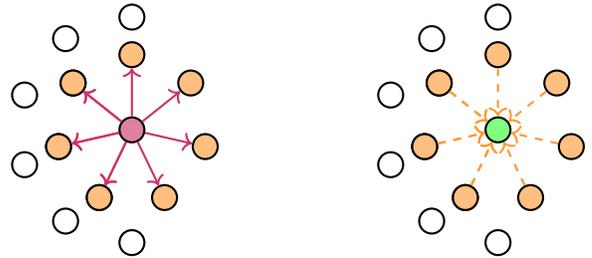

(Graph Sample and Aggregation) refer figure \ref{GraphSAGE} is a general inductive framework that leverages node feature information (e.g., text attributes) to efficiently generate node embeddings for previously unseen data. The node embedding is computed as:
\begin{align*}
\mathbf{h}_v^{(k)} = \sigma \left( \mathbf{W}^{(k)} \cdot \mathrm{aggregate}^{(k)} \left( \left\{ \mathbf{h}_u^{(k-1)}, \forall u \in \mathcal{N}(v) \right\} \right) \right)
\end{align*}
where:
    \begin{itemize}
        \item \(\mathbf{h}_v^{(k)}\) is the embedding of node \(v\) at layer \(k\)
        \item \(\sigma\) is an activation function (e.g., ReLU)
        \item \(\mathbf{W}^{(k)}\) is the weight matrix for layer \(k\)
        \item \(\mathrm{AGGREGATE}^{(k)}\) is an aggregation function (e.g., mean, LSTM) that combines the embeddings of the neighbors \(\mathcal{N}(v)\) of the node \(v\)
    \end{itemize}
\subsubsection{GCN}(Graph Convolutional Networks) \ref{GCN}utilize a convolutional operation on graphs to generate embeddings by aggregating information from a node's neighbors. The layer-wise propagation rule is defined as:
\begin{align*}
       \mathbf{H}^{(l+1)} = \sigma \left( \mathbf{D}^{-\frac{1}{2}} \mathbf{A} \mathbf{D}^{-\frac{1}{2}} \mathbf{H}^{(l)} \mathbf{W}^{(l)} \right) 
\end{align*}    
where:
    \begin{itemize}
        \item \(\mathbf{H}^{(l)}\) is the hidden representation at layer \(l\).
        \item \(\mathbf{A}\) is the adjacency matrix.
        \item \(\mathbf{D}\) is the degree matrix.
        \item \(\mathbf{W}^{(l)}\) is the weight matrix for layer \(l\).
        \item \(\sigma\) is an activation function (ReLU).
    \end{itemize}   
    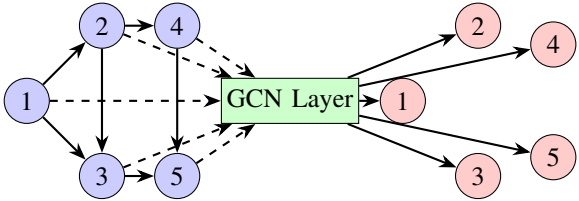
\begin{figure}[ht]
    \centering
    \begin{tikzpicture}[
      node/.style={circle, draw, fill=blue!20, minimum size=6mm, inner sep=0pt},
      edge/.style={-Stealth, thick},
      transform/.style={rectangle, draw, fill=green!20, minimum size=6mm, inner sep=2pt},
    ]
 
    \node[node] (v1) at (0,0) {1};
    \node[node] (v2) at (1,1) {2};
    \node[node] (v3) at (1,-1) {3};
    \node[node] (v4) at (2,1) {4};
    \node[node] (v5) at (2,-1) {5};

    \draw[edge] (v1) -- (v2);
    \draw[edge] (v1) -- (v3);
    \draw[edge] (v2) -- (v4);
    \draw[edge] (v3) -- (v5);
    \draw[edge] (v2) -- (v3);
    \draw[edge] (v4) -- (v5);
    \node[transform] (h1) at (3.5,0) {GCN Layer};
    \draw[edge, dashed] (v1) -- (h1);
    \draw[edge, dashed] (v2) -- (h1);
    \draw[edge, dashed] (v3) -- (h1);
    \draw[edge, dashed] (v4) -- (h1);
    \draw[edge, dashed] (v5) -- (h1);
    \node[node, fill=red!20] (o1) at (5,0) {1};
    \node[node, fill=red!20] (o2) at (6,1) {2};
    \node[node, fill=red!20] (o3) at (6,-1) {3};
    \node[node, fill=red!20] (o4) at (7,0.75) {4};
    \node[node, fill=red!20] (o5) at (7,-0.75) {5};
    \draw[edge] (h1) -- (o1);
    \draw[edge] (h1) -- (o2);
    \draw[edge] (h1) -- (o3);
    \draw[edge] (h1) -- (o4);
    \draw[edge] (h1) -- (o5);
    \end{tikzpicture}
    \caption{Illustration of GCN - The Graph aka adjacency matrix is sent into the GCN layer and then the embeddings of each node are produced as output}
    \label{GCN}
    \end{figure}

\subsection{Model Pipeline}
As a back-end of the model, we have the data flowing through the graph algorithms to predict the embeddings of the unseen nodes after learning the attributes of known nodes. Then, these embeddings are used as a vector representation of the song. The Euclidean distance between the corresponding songs is calculated and the sorted list of top 10 songs is recommended as the most similar songs for any given song.  

\begin{figure}
    \centering
    \includegraphics[width=1.0\linewidth]{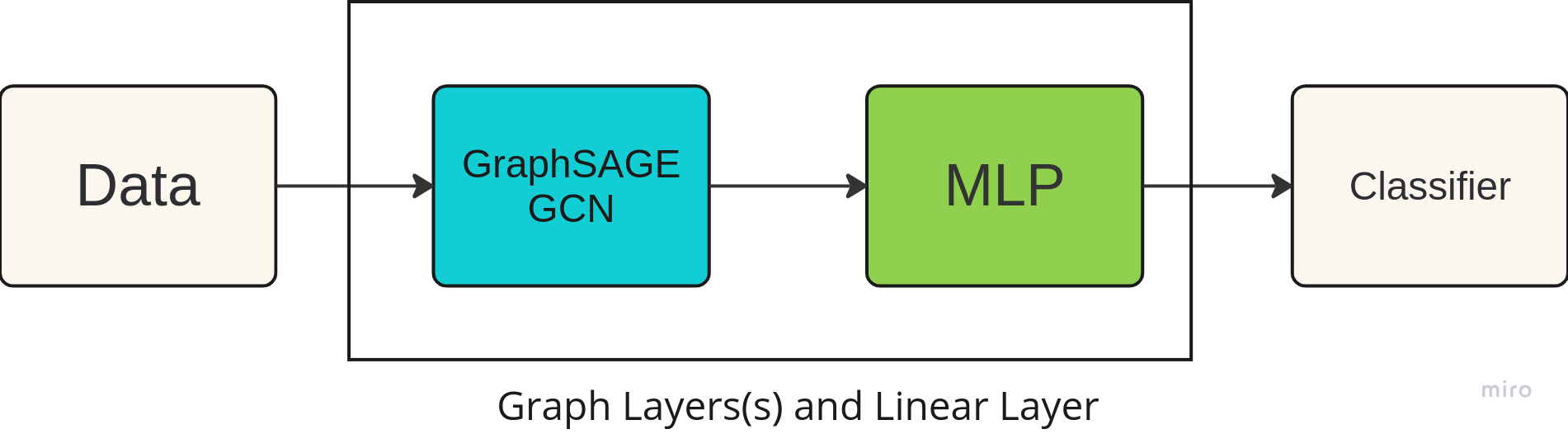}
    \caption{Algorithm Pipeline - the boxed structure refers to the models employed (Graph Sample \& Aggregation and Graph Convolutional Networks}
    \label{pipeline}
\end{figure}

Below is the full pipeline from start to finish involving learning MFCC features and MLP architecture producing a multi-class output (essentially a classification problem). \ref{pipeline}

\begin{figure}
    \centering
    \includegraphics[width=1.0\linewidth]{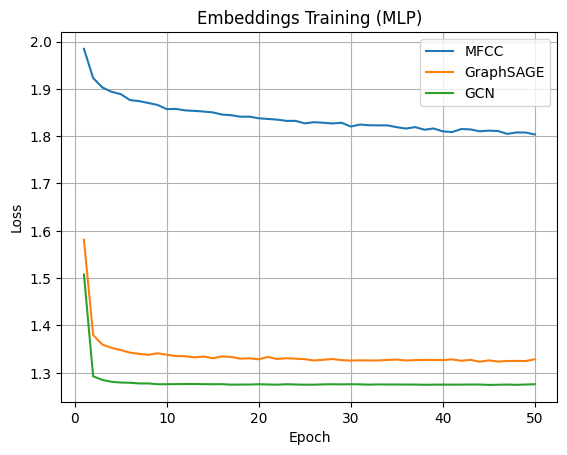}
    \caption{Training Plot of Plain MFCC, MFCC enriched with Graph Sample and Aggregation (GraphSAGE) and Graph Convolutional Networks (GCN)}
    \label{train_plot}
\end{figure}

\section{Experiments and Results}
\subsection{Experimental design}
To evaluate the performance of graph-based models in enhancing music genre recommendations, we conducted a series of experiments using a balanced dataset of 1000 songs from each of the 8 genres: Electronic, Experimental, Folk, Hip-Hop, Instrumental, International, Pop, and Rock. Each song was represented by a random 5-second window, and the Mel-Frequency Cepstral Coefficients (MFCC) features were extracted from these audio samples. The learning process of embeddings via the graph models and the MLP had the following hyperparameters shown in Table \ref{hyperparameters}

\begin{table}[ht]
\centering
\scriptsize
\caption{Hyperparameter Configurations}
\begin{tabular}{|p{2.8cm}|p{2.4cm}|p{2.4cm}|}
\hline
\textbf{Aspect} & \textbf{Parameter} & \textbf{Value} \\
\hline
\multirow{4}{*}{Embedding Learning} & Loss Function & Cross Entropy Loss \\
& Optimizer & Adam Optimization \\
& Learning Rate & 0.01 \\
& Model & GraphSAGE and GCN \\
\hline
\multirow{4}{*}{Genre Prediction} & Loss Function & Cross Entropy Loss \\
& Optimizer & Adam Optimization \\
& Learning Rate & 0.001 \\
& Model & 3 FC + Softmax \\
\hline
\end{tabular}
\label{hyperparameters}
\end{table}

The training plot graph of loss vs epochs (count = 50) tell us that GCN refined MFCC features had the least loss during training and it is also observed that it works well during testing as well. So, there is no overfitting here with regard to GCN. 
Additionally, the count of trainable parameters in each layer of the model pipeline is described as below. 
\begin{itemize}
    \item GraphSAGE MFCC $\rightarrow$ 3660 + MLP $\rightarrow$ 8360 = 12020    
    \item GCN MFCC $\rightarrow$ 1860 + MLP $\rightarrow$ 8360 = 10220    
    \item Plain MFCC = MLP = 8360
\end{itemize}

\subsection{Algorithmic Results}
After the embeddings (refined representation) for MFCC features were learned, using two graph models: GraphSAGE and Graph Convolutional Network (GCN), the predicted embeddings for an unseen song showed that the GraphSAGE model achieved an accuracy of 94\%, while the GCN model performed even better with an accuracy of 100\%. These results indicate that the embeddings produced and refined through these graph-based models are significantly more effective for music recommendation compared to the baseline accuracy of 29\%, which was obtained using the plain MFCC features generated by the librosa Python library \cite{librosa2025}. This is an assertion that graph-based model learning can be a useful tool for music recommendation systems.

The accuracy was calculated based on the below formula:

\begin{align*}
\Gamma = \frac{1}{N} \sum_{i=1}^{N} \left( \frac{1}{100} \sum_{j=1}^{100} \frac{R_{ij}}{10} \right)
\end{align*}
where:
\begin{itemize}
    \item \( \Gamma \) is the overall accuracy of the recommendations.
    \item \( N \) is the number of genres.
    \item \( i \) is the index for genres, ranging from \(1\) to \(N\).
    \item \( j \) is the index for the sounds within each genre, ranging from \(1\) to \(100\).
    \item \( R_{ij} \) is the number of correct recommendations for sound \( j \) in genre \( i \).
\end{itemize}

The below table shows the average accuracy recommendations for 100 songs chosen randomly for each of the 8 genres found for every embedding type that we have employed in the study (Plain MFCC embeddings, SAGE embeddings, GCN embeddings. Table \ref{recommendations}. It also provides a detailed comparison of three different music recommendation approaches—MFCC-based, GraphSAGE-based, and GCN-based—across eight music genres: Electronic, Experimental, Folk, Hip-Hop, Instrumental, International, Pop, and Rock. The results demonstrate a clear and consistent trend where graph-based methods (GraphSAGE and GCN) significantly outperform the plain MFCC features approach regarding accurate recommendations for each genre. 

\begin{table}[h]
\centering
\caption{Recommendation Algorithm Performance by Music Genre - Accuracy ($\Gamma$)}
\begin{tabular}{l|c|c|c}
\toprule
& \multicolumn{3}{c}{\texttt{Embedding Used}} \\
\midrule
\texttt{Music Genre} & \texttt{MFCC} & \texttt{GraphSAGE} & \texttt{GCN} \\
\midrule
Electronic & 28.00 & 100.00 & 100.00 \\
Experimental & 17.60 & 99.70 & 100.00 \\
Folk & 30.40 & 96.40 & 100.00 \\
Hip-Hop & 29.20 & 100.00 & 100.00 \\
Instrumental & 29.30 & 69.10 & 100.00 \\
International & 33.80 & 100.00 & 100.00 \\
Pop & 18.20 & 88.40 & 100.00 \\
Rock & 46.00 & 99.40 & 100.00 \\
\midrule
Avg. Accuracy ($\Gamma$) & 29.06 & 94.12 & 100.00 \\
\bottomrule
\end{tabular}
\label{recommendations}
\end{table}

For the Electronic, Experimental, Folk, Instrumental, and Pop genres, the plain MFCC features method achieves a paltry accurate recommendation, while both GraphSAGE and GCN achieve high scores and in fact sometimes score a perfect 10 for a particular genre (e.g, Pop for example), highlighting the substantial improvement graph embeddings offer. The GraphSAGE embeddings struggle to identify the `Instrumental' genre and have the lowest score among all other genres. This brings down the overall accuracy of the GraphSAGE model. But otherwise, the GraphSAGE model performs better for all the different genres scoring greater than at least 85\%.  The lowest score for the plain MFCC is obtained for Pop which could be attributed to the fact that the model is confused to differentiate between different genres. In general, the graph-based models show advanced capabilities in handling these more nuanced genres. GCN achieves perfect scores, further asserting its efficiency of it in understanding and recommending within any particular genre.

Overall, the data underscores the substantial benefits of integrating graph-based embeddings into music recommendation systems. GraphSAGE and GCN consistently enhance recommendation accuracy, with GCN occasionally showing a slight edge over GraphSAGE. These results affirm that graph-based approaches are crucial for capturing the complex and contextually rich relationships within music data, leading to more accurate and personalized music recommendations across a diverse array of genres. The potential reasons for this can be attributed to the fact that the aggregating method in GCN is simple (mean averaging) and is not complex (Long Short-Term Memory (LSTM) pooling. In addition to this, GCNs have parameter sharing that has been quite effectively adapted to this dataset. Moreover, the problem of overfitting is not a problem in the GCNs. GraphSAGE model did overfit and hence the drop in the accuracy is justified in the inference stage. Finally, the homogeneity of the dataset attributed to this as a simpler aggregation mechanism is more than sufficient than more complex aggregation methods as stated in the GraphSAGE model. 

The GitHub repository \href{https://github.com/bharani1990/CASM_Project}{here} contains the full example to reproduce the results obtained.

\section{Conclusion}
In this paper, we explored the integration of graph-based methods into music recommendation systems, focusing on enhancing genre-based recommendations through the use of Graph Convolutional Networks (GCN) and GraphSAGE. Our findings demonstrate the superiority of graph-based approaches over traditional methods relying solely on Mel-Frequency Cepstral Coefficients (MFCC). The results indicate that both GCN and GraphSAGE significantly outperform the MFCC-only approach across all tested genres, achieving near-perfect recommendation accuracies in many cases. By incorporating graph embeddings, we were able to capture complex relationships and contextual information often missed by traditional methods, leading to more accurate and relevant music recommendations. Both GCN and GraphSAGE showed scalability when applied to large music datasets, with GraphSAGE particularly designed to handle large-scale graphs efficiently by sampling and aggregating features from a fixed number of neighbors. The robustness of these approaches in handling diverse genres and providing highly personalized recommendations highlights their potential to enhance user satisfaction and engagement on music streaming platforms \cite{ying2018graph, wang2019kgat}.

Graph-based methods like GCN and GraphSAGE enhance recommendation accuracy but come with challenges, mainly in computational complexity and data requirements. GCN demands significant resources due to matrix multiplications for all nodes and their neighbors, which increases with graph size. GraphSAGE, while more scalable, still involves complex feature aggregation operations and can be resource-intensive for very large graphs. Effective graph-based recommendations rely on high-quality, comprehensive datasets that accurately represent music item relationships. Despite these challenges, the improvements in recommendation accuracy and the nuanced capture of music relationships highlight the transformative potential of GCN and GraphSAGE in music recommendation. Although Graph Transformers were initially considered, their use was deferred due to hardware limitations and the computational complexity of graph attention networks. Future work will include subjective analysis and explore the use of Graph Transformers.

\bibliographystyle{IEEEtran}
\bibliography{references}

\begin{thebibliography}{10}
\providecommand{\url}[1]{#1}
\csname url@samestyle\endcsname
\providecommand{\newblock}{\relax}
\providecommand{\bibinfo}[2]{#2}
\providecommand{\BIBentrySTDinterwordspacing}{\spaceskip=0pt\relax}
\providecommand{\BIBentryALTinterwordstretchfactor}{4}
\providecommand{\BIBentryALTinterwordspacing}{\spaceskip=\fontdimen2\font plus
\BIBentryALTinterwordstretchfactor\fontdimen3\font minus
  \fontdimen4\font\relax}
\providecommand{\BIBforeignlanguage}[2]{{%
\expandafter\ifx\csname l@#1\endcsname\relax
\typeout{** WARNING: IEEEtran.bst: No hyphenation pattern has been}%
\typeout{** loaded for the language `#1'. Using the pattern for}%
\typeout{** the default language instead.}%
\else
\language=\csname l@#1\endcsname
\fi
#2}}
\providecommand{\BIBdecl}{\relax}
\BIBdecl

\bibitem{nam2019deep}
J.~Nam, K.~Choi, J.~Lee, S.-Y. Chou, and Y.-H. Yang, ``Deep learning for
  audio-based music classification and tagging: Teaching computers to
  distinguish rock from bach,'' \emph{IEEE Signal Processing Magazine},
  vol.~36, no.~1, pp. 41--51, 2019.

\bibitem{wang2021graph}
\BIBentryALTinterwordspacing
S.~Wang, L.~Hu, Y.~Wang, X.~He, Q.~Z. Sheng, M.~A. Orgun, L.~Cao, F.~Ricci, and
  P.~S. Yu, ``Graph learning based recommender systems: A review,'' 2021.
  [Online]. Available: \url{https://arxiv.org/abs/2105.06339}
\BIBentrySTDinterwordspacing

\bibitem{sakurai2022graph}
\BIBentryALTinterwordspacing
K.~Sakurai, R.~Togo, T.~Ogawa, and M.~Haseyama, ``Controllable music playlist
  generation based on knowledge graph and reinforcement learning,''
  \emph{Sensors}, vol.~22, no.~10, 2022. [Online]. Available:
  \url{https://www.mdpi.com/1424-8220/22/10/3722}
\BIBentrySTDinterwordspacing

\bibitem{hamilton2017inductive}
\BIBentryALTinterwordspacing
W.~L. Hamilton, R.~Ying, and J.~Leskovec, ``Inductive representation learning
  on large graphs,'' 2018. [Online]. Available:
  \url{https://arxiv.org/abs/1706.02216}
\BIBentrySTDinterwordspacing

\bibitem{velivckovic2018graph}
\BIBentryALTinterwordspacing
P.~Veličković, G.~Cucurull, A.~Casanova, A.~Romero, P.~Liò, and Y.~Bengio,
  ``Graph attention networks,'' 2018. [Online]. Available:
  \url{https://arxiv.org/abs/1710.10903}
\BIBentrySTDinterwordspacing

\bibitem{kipf2016graph}
\BIBentryALTinterwordspacing
T.~N. Kipf and M.~Welling, ``Semi-supervised classification with graph
  convolutional networks,'' 2017. [Online]. Available:
  \url{https://arxiv.org/abs/1609.02907}
\BIBentrySTDinterwordspacing

\bibitem{mfcc}
Z.~K. Abdul and A.~K. Al-Talabani, ``Mel frequency cepstral coefficient and its
  applications: A review,'' \emph{IEEE Access}, vol.~10, pp.
  122\,136--122\,158, 2022.

\bibitem{fma2016}
\BIBentryALTinterwordspacing
M.~Defferrard, K.~Benzi, P.~Vandergheynst, and X.~Bresson, ``Fma: A dataset for
  music analysis,'' 2017. [Online]. Available:
  \url{https://arxiv.org/abs/1612.01840}
\BIBentrySTDinterwordspacing

\bibitem{librosa2025}
\BIBentryALTinterwordspacing
B.~McFee~et al, ``librosa/librosa: 0.11.0,'' Mar. 2025. [Online]. Available:
  \url{https://doi.org/10.5281/zenodo.15006942}
\BIBentrySTDinterwordspacing

\bibitem{ying2018graph}
\BIBentryALTinterwordspacing
R.~Ying, R.~He, K.~Chen, P.~Eksombatchai, W.~L. Hamilton, and J.~Leskovec,
  ``Graph convolutional neural networks for web-scale recommender systems,'' in
  \emph{Proceedings of the 24th ACM SIGKDD International Conference on
  Knowledge Discovery and Data Mining}.\hskip 1em plus 0.5em minus 0.4em\relax
  ACM, Jul. 2018, p. 974–983. [Online]. Available:
  \url{http://dx.doi.org/10.1145/3219819.3219890}
\BIBentrySTDinterwordspacing

\bibitem{wang2019kgat}
\BIBentryALTinterwordspacing
X.~Wang, X.~He, Y.~Cao, M.~Liu, and T.-S. Chua, ``Kgat: Knowledge graph
  attention network for recommendation,'' in \emph{Proceedings of the 25th ACM
  SIGKDD International Conference on Knowledge Discovery and Data
  Mining}.\hskip 1em plus 0.5em minus 0.4em\relax ACM, Jul. 2019, p. 950–958.
  [Online]. Available: \url{http://dx.doi.org/10.1145/3292500.3330989}
\BIBentrySTDinterwordspacing

\end{thebibliography}

\vspace{12pt}
\color{red}

\end{document}